\documentstyle[12pt]{article}
\def \a {\alpha}
\def \b {\beta}
\def \d {\partial}
\def \s {\sigma}
\def \lra{\longrightarrow}
\def \e {\epsilon}
\def \t {\theta}
\def \T {\Theta}

\def \P {\Phi}
\def \de {\delta}

\def \be {\begin{equation}}
\def \ee {\end{equation}}
\def \l {\lambda}
\def \w {\omega}
\begin{document}
\begin{titlepage}
\begin{flushright}
SU-4240-656\\
IMSc 97/02/03 \\
Feb 97
\end{flushright}

\centerline{\bf Hamiltonian formulation of Quantum Hall Skyrmions 
with Hopf term}

\begin{center}
{\bf B. Chakraborty$^*$}\\
{\em S. N. Bose National Center for Basic Sciences\\
Block JD, Sector III, Salt lake, Calcutta 700 091, India}\\
and\\
{\bf T. R. Govindarajan$^\dagger$}\\
{Dept of Physics, Syracuse University\\
Syracuse, NY 13244-1130, USA}\\
and\\
The Institute of Mathematical Sciences$^{\dagger\dagger}$\\
Chennai 600 113, India\\
\end{center}
\vskip1cm
\noindent{\bf Abstract}

We study the nonrelativistic nonlinear sigma model with Hopf term in this
paper. This is an important issue because of its relation to the 
currently interesting studies in skyrmions in quantum Hall systems.
We perform the Hamiltonian analysis of this system in $CP^1$ 
variables. When the coefficient of the Hopf term becomes zero
we get the Landau-Lifshitz description of the ferromagnets. The addition
of Hopf term dramatically alters the Hamiltonian analysis. 
The spin algebra is modified giving a new structure and interpretation
to the system. We point out momentum and angular momentum 
generators and new features they bring in to the system.

\vfill
\hrule
\vskip1mm
\noindent{\em * biswajit@bose.ernet.in}\\
\noindent{\em $\dagger$govind@suhep.phy.syr.edu; trg@imsc.ernet.in}
\noindent{\em $\dagger\dagger$ permenant address}
\vskip.5cm
\end{titlepage}

\noindent{\bf 1. Introduction}

It is well known that static and dynamical properties of 
ferromagnets are captured by the Landau-Lifshitz(LL) 
evolution equations\cite{polyakov}. These equations can be obtained from LL
Hamiltonian which is the continuum limit of the Heisenberg 
spin chain exploiting the modified Poisson brackets 
among the magnetisation fields. A lagrangian description is 
possible through the non relativistic non-linear sigma model(NLSM).
This model has captured attention recently \cite{sondhi}
for the description of 
novel excitations known as skyrmions in a suitable 
quantum Hall regime. Conventional quantum Hall regime is one
in which the magnetic field is strong enough that physical
properties are robust against changes coming from mixing of Landau 
levels through interactions. Essentially one assumes that in this 
regime the Landau level seperation $\hbar \w_c$ is larger than 
any other energy scale in the system. For free electrons in
magnetic field the Zeeman splitting $(g\mu_B B)$ is of the same order 
as the Landau level seperation\cite{qhf}. However electrons in conduction band 
have renormalised mass and g-factor. In the typical case of GaAs
the the mass gets enhanced by a factor of 20 and the Zeeman splitting is 
reduced by a factor of 4\cite{qhf1}. In such a case we have a Quantum Hall
ferromagnet described by a nonrelativistic non-linear sigma model. It
is well known that in such a system there are solitons  
arising from purely topological properties of the configuration
space\cite{skyrme}. It is an interesting question to ask whether the spin and 
statistics of these solitons can also obtained by topological
considerations. Such a question has been studied extensively in 
relativistic NLSM and it is well known that 
the spin and statistics can be obtained through
a well known Hopf term\cite{wil}. Recently Chandar etal\cite{trg}
have studied the many electron system (2DEG) in a 
quantum Hall ferromagnetic regime 
and obtained a nonzero contribution to the Hopf term. This term 
turns out to be $\propto \nu$, the filling fraction. 
The role of this term has never been analysed in a Hamiltonian
framework in a nonrelativistic context. On the other hand it is
an important analysis to be performed since the symplectic 
structure of the system is far different from the relativistic
context. The hamiltonian analysis has been studied earliar without the Hopf
term\cite{chaks}.
 
In this paper, we study the implications of the addition of the 
Hopf term to the nonrelativistic NLSM.
by carrying out
the Hamiltonian analysis in a gauge independent manner
\cite{hanson,ban}.Recently this gauge independent
scheme has been used in various models involving
Chern-Simons(CS) term \cite{ban,banbis,biswa} revealing the
existence of fractional spin. This has certain advantages over
gauge fixed approach. For example, the relevant symmetry is
manifested right at the level of transformation properties of the
basic fields, even if they are gauge variant.On the other hand
the transformation property of these fields gets affected by the
gauge fixing condition used \cite{hanson,biswa}.
The transformation property of the gauge invariant objects
however remains unaffected by gauge fixing.The underlying
symmetry group is therefore uncovered only at the level of gauge
invariant variables. Besides,the symplectic structures, given by
the set Dirac brackets (DB) usually becomes more complicated so
that subsequent quantization by elevating DB to quantum
commutators may have serious operator ordering ambiguities \cite{hanson,
archan}.
Indeed,as we shall see that the symplectic structure in our
model in the gauge independent scheme is complicated enough, let
alone in the gauge fixed scheme, so that we are forced to
restrict our analysis to the classical level only. In sec 2 we 
describe the model. In sec 3. we perform the Hamiltonian analysis 
of the model followed by a discussion of the system in sec 4. 
   
\noindent{\bf 2. The model}

As explained in the introduction in the long wavelength limit the excitations 
near  ferromagnetic ground state can be described by a non-relativistic 
NLSM\cite{nlsm}. As mentioned earliar this provides  a good model 
in some quantum Hall systems where the g-factor is small\cite{sondhi}.
In this case there are  excitations which are  the solitons of the NLSM.
This model is described by a field ${\bf n}(x,y,t)$ satisfying the 
condition ${\bf n}^2~=~1$\cite{polyakov,wil}. The configuration
space $Q$ is given by the set of maps $\{{\bf n}\}$, 
\be {\bf n}~:~ R^2 \lra S^2
\ee
For the finite energy configurations to exist in this model we need to impose
the condition that ${\bf n}(x,y,t)~~~{\buildrel {x,y \rightarrow \infty}
\over \lra}~~~ Const$.
With this condition we have a compactified $R^2$ which is the same as an
$S^2$ so that the configuration space is  given by
\be {\bf n}~:~ S^2\lra S^2
\ee
The configuration space splits into disjoint union of path connected spaces
$Q_N$ each labelled by an integer $N$ known as soliton number. This easily
follows from the fact that\cite{homotopy}
\be \pi_0(Q)~=~ \pi_2(S^2)~=~ {\cal Z}
\ee
The soliton number N can be obtained through a topological conservation
law for the current $j^\mu,$ given by\cite{skyrme}
\be
j^\mu~=~{1\over 8\pi}\e^{\mu\nu\l}{\bf n}\cdot {\d}_\nu{\bf n}\times {\d}_\l{\bf n}
\label{topo}
\ee
$j_0$ of this current is the soliton density for the map.
\be
N~=\int d^2x~j_0(x)~~~~\label{soliton}
\ee
Note that the conservation of the current $j^\mu$ in eq.[\ref{topo}] holds 
without reference to 
the equations of motion. Using the conservation of the topological current
we can write 
\be j^\mu~=~{1\over 2\pi}\e^{\mu\nu\l}\d_\nu a_\l \label{jmu}
\ee
where $a$ is one-form obtained by pulling back onto the space-time
the magnetic monopole connection of Dirac\cite{dirac}. 
If we parametrize $n$ through $\T$ and $\P$,
the polar cordinates for $S^2$ then $j$ can be written as mentioned earliar
in eq.[\ref{jmu}] as the dual of 
a two-form $da$. Explicitely, $a_\mu$ is given by:
\be
a_\mu~=~{1\over 2}(\pm 1 ~-~\cos \T)\d_\mu\P
\ee
It is also well known that in all the soliton number\cite{wil,homotopy} 
sectors the fundamental group is nontrivial. This follows from
\be
\pi_1(Q)~=~\pi_3(S^2)~=~{\cal Z}
\ee
This implies that the loops based at any point in the configuration
space Q falls into separate homotopy classes labelled by an integer.
This integer is obtained by the Hopf-term\cite{wil,homotopy}
\be
{\cal H}~=~ \int d^3x j^\mu a_\mu. \label{hopf}
\ee
where $a$ is defined through eq.[\ref{jmu}]. 

It is well known that the introduction the Hopf-term in the action 
of the relativistic NLSM has the effect of changing the spin 
and statistics of the soliton\cite{wil,dimitra}. 
This will be so in the antiferromagnetic
system. Here one is studying the effect of this term in the 
nonrelativistic NLSM. One is interested in studying the effect of 
adding this term to the ferromagnetic system which leads to nonrelativistic 
NLSM with Hopf term. 
As was pointed out specifically in the quantum Hall context
it has been shown \cite{trg} that this term gets generated with 
a coefficient determined by the filling fraction. They have also shown
in path integral formalism that statistics of skyrmion is 
determined by the same coefficient. In this paper we would like to study
in Hamiltonian analysis the role of the same term. For this purpose it is 
advantageous to use $CP^1$
variables to describe the NLSM instead of $S^2$ ones.
Firstly the  NLSM defined in
terms of spin variables ${\bf n}$  has a Dirac singularity and 
this will disappear in the $CP^1$ description\cite{zee}. This will help 
in defining the generators of the symmetry transformations 
unambiguosly\cite{chaks}. This singularity is related to the 
the fact that the U(1) principal bundle over $S^2$ is nontrivial
and there is no global section\cite{maths}. Secondly the Hopf-term 
in the spin variables is a non-local term in the action.
This can be avoided by  enhancing the degrees of freedom and describing 
the same as a local action. This can be easily seen by recalling 
that $CP^1$ manifold is described by $Z~=~\pmatrix {Z_1&Z_2}$
with $Z^\dagger Z~=~1$ and the identification $Z~\sim~e^{i\t}Z$.
where $e^{i\t}$ is an U(1) element. The one-form $a_{\mu}$ 
can be obtained by geometrical considerations \cite{chaks}, to get
\be
a_\mu ~=~ -iZ^\dagger \d_\mu Z \label{jmu1}
\ee
The topological current 
in $CP^1$ variables can therefore easily be seen to be given by
\be
j^\mu~=~{-i\over 2\pi}\e^{\mu\nu\l}\d_\nu Z^\dagger \d_\l Z \label{topo1}
\ee
The Hopf-term becomes,
\be
{\cal H}~=~ -{1\over 2\pi}\e^{\mu\nu\l}\int d^3x Z^\dagger \d_\mu Z
      \d_\nu Z^\dagger \d_\l Z \label{hopf1}
\ee
Clearly this expression\cite{zee} for the Hopf-term is local unlike that in
spin variables given in eq.[\ref{hopf}]. The locality is achieved through 
a gauge invariance and it is the same expression whether one is considering 
nonrelativistic or relativistic NLSM. The action corresponding to the 
nonrelativistic NLSM can be written as 
\be S~=~-\int d^3x\{{i\over 2}Z^\dagger {\stackrel{\leftrightarrow}{\d_0}}Z~
+~A_0(Z^\dagger Z-1)~+~
|D_iZ|^2\} ~+~\t {\cal H} \label{action}
\ee

Here while considering the action [eq.\ref{action}] 
we find it convenient to enlarge the
phase space and intoduce a new gauge field $a_\mu$. This $a_\mu$ gauge field
should not be confused with the one introduced earliar even though
equations of motion will relate them. We couple this 
gauge field to the topological current and add only the Chern Simons
term for this gauge field\cite{trgrs}.
\be 
S~=~ S_0 ~+~\t \e_{\mu\nu\l} \int d^3x \{2ia_\mu \d_\nu Z^\dagger \d_\l Z
~+~a_\mu \d_\nu \a_\l \} \label{action1}
\ee 
This action in eq.[\ref{action1}] is same as the previous one 
in eq.[\ref{action}] when we use the equations of motion
for the $a_\mu$. 

In this paper we will deal with the action eq.[\ref{action1}] and 
carry out Hamiltonian 
analysis. We will find that the Hopf term dramatically changes the
content of the theory by modifying the spin algebra. We will follow the 
Faddeev-Jackiw \cite{fj} symplectic analysis for the first order Lagrangians.
This will ease our constraint analysis and help in obtaining the new Dirac-
Brackets. 

\noindent{\bf 3. Hamiltonian Analysis}

In this section we shall perform Hamiltonian analysis of the model
[\ref{action1}]. The Lagrangian corresponding to [\ref{action}] is given by 
\be
L=\int d^2x\{-{i\over 2}Z^\dagger {\stackrel{\leftrightarrow}{\d_0}}Z
-A_0(Z^\dagger Z -1)-
|D_iZ|^2+\t \e^{\mu\nu\l}a_\mu(2i \d_\nu Z^\dagger \d_\l Z~ +\d_\nu a_\l)\}
\label{lagrangian}
\ee
Here we would like to point out that the covariant derivative $D$
in eq.(\ref{lagrangian}) involves only the background 
gauge-field $A$ and is given by 
$D_i~=~\d_i~-~iA_i$ and does not involve the other gauge field 
$a_\mu$. The configuration space variables are $~[a_0(x),a_i(x),A_i(x),A_0(x),
Z_\a(x),Z_\a^*(x)]$. However the role of $A_0$ and $a_0(x)$ are basically
that of lagrange multipliers. They enforce the constraints 
\be G_1(x)\equiv Z^\dagger (x)Z(x)~-~1\approx~0 \label{cnstrnt1}
\ee
\be G_2(x)\equiv \e_{ij}(i\d_iZ^\dagger(x)\d_jZ(x)~+~\d_ia_j(x))\approx 0
\label{cnstrnt2}
\ee
respectively.

The conjugate momenta to $(A_i(x),a_i(x),Z_\a(x),Z^*_\a(x))$ are given by
\be 
\Pi_i(x)\equiv {\de L\over \de{\dot A}_i(x)}~=~0\label{pc1}
\ee
\be\pi_i(x)\equiv {\de L\over \de {\dot a}_i(x)}~=~\t\e_{ij}a_j
\label{pc2}\ee
\be {\cal P}_\a(x)\equiv {\de L \over \de {\dot Z}_\a(x)}~=~
-{i\over 2}Z^*_\a(x)~+~2i\t\e_{ij}a_i\d_j Z^*_\a(x)
\label{pc3}\ee
\be {\cal P}_a^*(x)\equiv {\de L\over \de{\dot Z}_\a^*(x)}~=~
{i\over 2}Z_\a(x)~-~2i\t\e_{ij}a_i(x)\d_jZ_\a(x)
\label{pc4}\ee
These are the set of primary constraints of the model. Since the model is 
first order in time-derivative the Hamiltonian can be readily obtained as 
\be H_c~=~ \int d^2x[|D_iZ|^2~+~A_0(x)G_1(x)~-~2\t a_0(x)G_2(x)]
\label{hamilton}
\ee
By requiring the constancy of primary constraints (\ref{pc1}) 
we get the following secondary constraints:
\be
C_i(x)~\equiv A_i(x)~+~iZ^\dagger(x)\d_i Z(x)\approx 0
\label{sc1}\ee
Inview of this strong equality $A_i$ ceases to be an independent
degree of freedom.
Clearly this constraint is conjugate to the constraint(\ref{pc1}) 
and can be strongly implemented by the Dirac bracket 
\be \{A_i(x),\Pi_j(y)\}~=~0
\ee
With this $A_i$ becomes,
\be
A_i(x)~\approx ~-iZ^\dagger \d_i Z(x)\label{AI}
\ee
The CS gauge-field also has interestingly the same form:
\be
a_i\approx -iZ^\dagger\d_iZ\label{ai}.
\ee
Thus both the gauge fields agree modulo gauge transformations on the 
constraint surface. However they are not identical and must be treated as 
distinct.

We are thus left with constraints (\ref{pc2}-\ref{pc4}).
These are the second class constraints of the model
as the Lagrangian (\ref{lagrangian}) is linear in ``velocities'' of the 
associated variables{\cite{fj}}.The constraints imposed by the 
lagrange multipliers $A_0$ and $a_0$ are expected to be gauss-law constraints
as in electrodynamics. Since gauss's law constraint is a gauge 
transformation generator we expect these to be first-class constraints. 
Note in our model we have $U(1)\times U(1)$ gauge symmetry correesponding to 
the two gauge fields $A_i(x)$ and $a_i(x)$.
We will now demostrate the fact that (\ref{cnstrnt1}) and (\ref{cnstrnt2}) 
are indeed first-class by explicite calculation. Thus to begin with, we 
have to invert the matrix formed by the poisson brackets of the constraints
(\ref{pc2}),(\ref{pc3}) and (\ref{pc4}). 
\be
\chi_i(x)\equiv \pi_i(x)~-~\t\e_{ij}a_j(x)~\approx~0 \label{pc2'}
\ee
\be
C^1_\a(x)\equiv {\cal P}_\a(x)~+~{i\over 2}Z^*_\a(x)~-~2i\t\e_{ij}a_i(x)
\d_jZ^*_\a(x)~\approx ~0 \label{pc3'}
\ee
\be
C^2_\a(x)\equiv (C^1_\a(x))^* \label{pc4'}
\ee
In order to obtain Dirac brackets we have to thus invert a $6\times 6$ matrix of 
the poisson brackets of constraints. We can simplify the procedure by 
doing the inversion in parts following Hanson etal\cite{hanson}.
Specifically note that pair of constraints in eq.(\ref{pc2'}) do not involve 
fields from matter sector. So we can readily implement this pair of 
constraints strongly by the Faddeev-Jackiw scheme\cite{biswa,archan,fj}
to yield the Dirac bracket:
\be
\{a_i(x),a_j(x)\}^{DB}~=~{\e_{ij}\over 2\theta}\de (x-y)
\label{gauge}\ee
However this is not the final DB in the gauge field sector. This, along
with the basic Poisson brackets $\{Z_\a ,Z_\b\}$ and $\{Z_\a ,a_i\}$
will undergo a further modification as we  
implement the other four constraints (\ref{pc3'}) and (\ref{pc4'}) 
strongly.This is because these constraints involve both $Z$ and
$a_i$ fields. To this end, we calculate the 
following matrix elements of Poisson bracket matrix (where the use of the 
Dirac bracket in eq.(\ref{gauge}) has been made).
\be
C^{11}_{\a\b}(x,y)~\equiv~\{C^1_\a(x),C^1_\b(y)\}~=~iR_{\a\b}(x)\de(x-y)
\label{c11}\ee
\be
C^{22}_{\a\b}(x,y)~\equiv~\{C^2_\a(x),C^2_\b(y)\}~=~iS_{\a\b}(x)\de(x-y)
\label{c22}\ee
\be
C^{12}_{\a\b}(x,y)~\equiv~\{C^1_\a(x),C^2_\b(y)\}~=~iT_{\a\b}(x)\de(x-y)
\label{c12}\ee
where 
\be
R_{\a\b}~=~-S^*_{\a\b}~=~2i\t\e_{\a\b}\nabla Z_1^*\times\nabla Z_2^*
\label{rab}\ee
\be
T_{\a\b}~=~T^*_{\b\a}~=~(\de_{\a\b}-2i\t\nabla Z_\a^*\times\nabla Z_\b)
\label{tab}\ee
The final matrix formed by the constraints (\ref{pc3'}) 
and (\ref{pc4'}) is therefore given
by
\be
C~=~\pmatrix {C^{11}(x,y)&C^{12}(x,y)\cr C^{21}(x,y)&C^{22}(x,y)}
\label{cab}\ee
The inverse of the matrix can readily be calculated:
\be
C^{-1}(x,y)~=~{-i\over {\cal D}}\pmatrix {0&-S_{12}&T_{22}&-T_{21}\cr
S_{12}&0&-T_{12}&T_{11}\cr -T_{22}&T_{12}&0&-R_{12}\cr T_{21}&-T_{11}&
R_{12}&0}\de (x-y)
\label{c-1ab}\ee
where 
\be
{\cal D}~=~R_{12}S_{12}~-~det T
\ee
We can simplify the expression for ${\cal D}$. At the end we get 
\be
{\cal D}~=~-(1~+~2\t B)
\label{cald}\ee
where $B~\equiv \e_{ij}\d_iA_j=~-i\nabla Z^\dagger \times \nabla Z$ is the 
magnetic field corresponding to the gauge field $A_i$. 
Upto a factor this is also the topological density
$j_0$. In order to make 
sense we have to impose the condition ${\cal D} \neq 0$ or equivalently
$B~\neq~-{1\over 2\t}$
The final Dirac brackets in the matter sector can be written as:
\be
\pmatrix {\{Z_1(x),Z_1(y)\}&\{Z_1(x),Z_2(y)\}\cr\{Z_2(x),Z_1(y)\}&
\{Z_2(x),Z_2(y)\}}~=~-{i\over {\cal D}}\pmatrix {0&-S_{12}\cr S_{12}&0}
\de (x-y)\label{matter}
\ee
\be
\pmatrix{\{Z_1(x),Z_1^*(y)\}&\{Z_1(x),Z_2^*(y)\}\cr\{Z_2(x),Z_1^*(y)\}&
\{Z_2(x),Z_2^*(y)\}}~=~-{i\over {\cal D}}\pmatrix {T_{22}&-T_{21}\cr
-T_{12}&T_{11}}\de (x-y)
\label{matter1}\ee
In addition to the above we have to obtain the Dirac brackets of the matter
sector with gauge fields. They can be worked out to be:
\be
\pmatrix { \{Z_1(x),a_k(y)\}\cr \{Z_2(x),a_k(y)\}\cr \{Z_1^*(x),a_k(y)\}
\cr \{Z_2^*(x),a_k(y)\}}~=~{1\over {\cal D}}
\pmatrix {-T_{22}&T_{21}&0&-S_{12}\cr T_{12}&-T_{11}&S_{12}&0\cr
0&R_{12}&-T_{22}&T_{12}\cr -R_{12}&0&T_{21}&-T_{11}}\pmatrix {\d_kZ_1\cr
\d_kZ_2\cr \d_kZ_1^* \cr \d_kZ_2^*}\de(x-y)
\label{gagmat}\ee
At this stage one can prove the following useful identities: 
\be
T_{11}\d_kZ_2~-~T_{12}\d_kZ_1~-~S_{12}\d_kZ^*_1~=~\d_kZ_2;~
T_{22}\d_kZ_1~-~T_{21}\d_kZ_2~+~S_{12}\d_kZ^*_2~=~\d_kZ_1\label{iden}
\ee
Using these identities the brackets from the mixed sector 
eq.(\ref{gagmat}) can be easily simplified to:
\be 
\{a_k(x),Z(y)\}~=~{1\over {\cal D}}\d_kZ~\de (x-y) \label{gagmat1}
\ee
Now we must go back and rework the new Dirac brackets between the 
gauge fields $a_i$. This turns out to be
\be
\{a_i(x),a_j(y)\}^{DB}~=~-{\e_{ij}\over 2\t{\cal D}}\de(x-y)\label{newdb}
\ee
We thus have at our disposal the Dirac brackets among all the field
variables. They are given by the equations (\ref{matter}) 
(\ref{matter1}), (\ref{gagmat1}) and (\ref{newdb}). Armed with these 
we can check that the constraints $G_1(x)$ and $G_2(x)$ generate
the gauge transformations.

Let us consider the constraint $G_1$ first. It is easy to check that 
\be
\int d^2y f(y)\{Z_1(x),G_1(y)\}~=~ {if(x)\over {\cal D}}(-T_{22}Z_1~+~T_{21}Z_2~
+~S_{12}Z_2^*)
\ee
where $f(x)$ is an arbitrary function with finite support.
Using the explicite forms of $T,S$ and ${\cal D}$ we can show that 
\be 
\de Z_\a(x)~=~\int d^2y f(y)\{Z_\a(x),G_1(y)\}\approx ~if(x)Z_\a(x)
\ee
It is not difficult to show that 
\be 
\{a_k(x),G_1(y)\}~\approx~0,~~~ and~~~ \{Z_\a(x), G_2(y)\}~\approx~0
\ee
This establishes that the constraint $G_1(x)$ has no effect on the gauge field
sector and similarly  $G_2(x)$ does not effect any 
transformation in the matter sector. We will now establish that 
it is exactly $2\t G_2(x)$ 
that generates the second $U(1)$ gauge transformation.
For this purpose note that $G_2(x)~=~b(x)-B(x)$ where $b(x)~=~\e_{ij}
\d_ia_j$ and $B(x)~=~\e_{ij}\d_i A_j$ are the `magnetic fields'
of the gauge potentials $a$ and $A$ respectively. Simple algebra leads to 
\be
\int d^2y f(y)\{a_i(x),B(y)\}~=~-{B\over {\cal D}}(\d_if);
\int d^2y f(y)\{a_i(x),b(y)\}~=~{1\over 2\t {\cal D}}(\d_if)
\ee
From this and the Dirac brackets of the gauge field 
sector and eq(\ref{gagmat1}),
we can arrive at the following relations using the
explicite form of ${\cal D}$ 
\be
\int d^2y f(y)\{a_i(x),{G_2}(y)\}\approx~{1\over 2\t}\d_if(x)
\ee
establishing our claim. It is easy to prove that these two generators
have zero Dirac brackets among themselves.

To proceed further with the analysis we consider the spin variables
$n_a(x)$ mentioned in the section 2. They can be obtained from the 
$CP^1$ variables by using the Hopf map:
\be 
n_a(x)~=~ Z^\dagger \s_a Z
\ee
We can obtain using eqs.({\ref{matter}) and ({\ref{matter1}) that 
\be
\{n_a(x),B(y)\}\approx~ {1\over {\cal D}}\nabla \de(x-y)\times \nabla n_a(x)
\ee
Given this we will now consider the modifications for the spin algebra
that is obtained by the introduction of the Hopf term in the action.

Using the Hopf map we can write down the components of spin variable
${\bf n}$ as:
\be n_1~=~Z_1Z_2^*~+~Z_1^*Z_2~;~n_2~=~i(Z_1Z_2^*~-~Z_1^*Z_2)~;~
n_3~=~|Z_1|^2~-~|Z_2|^2.
\ee
For convenience let us consider the Dirac bracket between $n_1$ and $n_2$.
This is given by, using eqs.(\ref{matter}) and (\ref{matter1}):
\be
\{n_1(x),n_2(y)\}~={1\over {\cal D}}[2n_3(x)~-~4i\t{\cal M}]\de(x-y)
\ee
where ${\cal M}$ is given by 
\be
{\cal M}~=~|Z_1|^2\nabla Z_1^*
\times \nabla Z_1~-~|Z_2|^2\nabla Z_2^*\times \nabla Z_2~
+(~Z_1Z_2\nabla Z_1^*\times \nabla Z_2^*~-c.c.)
%-~Z_1^*Z_2^*\nabla Z_1\times \nabla Z_2
\ee
One can easily see that $\t$ dependent term in the above is gauge 
invariant and hence we can evaluate it in a particular gauge. Following 
\cite{chaks},
we choose $Z_1~=~Z_1^*$ and find the term vanishes. Similarly we get the 
same result for the other choice $Z_2~=~Z_2^*$.
We therefore find that 
\be 
\{n_1(x),n_2(y)\}~=~{2 \over {\cal D}}n_3(x)\de(x-y)
\ee
It is easy to write this covariantly as:
\be 
\{n_a(x),n_b(y)\}~=~{2\over {\cal D}}\e_{abc}n_c(x)\de(x-y).
\ee
Note that in contrast to the NLSM without Hopf term we have an extra 
factor  ${\cal D}$ in the denominator. In view of the
nonvanishing bracket $\{n,B\}$ the conventional spin algebra
cannot be restored even if  we redefine a new renormalised 
spin variable ${\bf \tilde n}~=~{\cal D}{\bf n}$.

\noindent{\bf 4. Global generators and Discussions}

In this paper we have considered the Hamiltonian analysis of the non-
relativistic non linear sigma model with the topological Hopf term 
added. This model as was pointed out is quite relevant when we consider 
polarised quantum Hall system with non zero g-factor. As was pointed 
out in the literature there are solitons in this model and the 
introduction of the Hopf term was expected to alter the statistics
of these skyrmions. We have chosen to work with $CP^1$ variables
in order to avoid introduction of nonlocal terms which naturally
arise when we consider Hopf terms.
True to the expectations, the introduction of the 
term changes the symplectic structure defined by the system. 
Interestingly a modified `spin algebra' was obtained. In this
modification the topological charge density plays an important role.
Whenever ${\cal D}=0$ the spin algebra breaks down.
We have obtained the generators of the gauge transformations.
We verified that these generators produce appropriate 
transformations. 

For completenesss we can easily work out the other global symmetry 
generators also. For example, the Hamiltonian as time translations 
generator is given by 
\be
H~=~\int d^2x \{ |D_iZ|^2~+~A_0(x)G_1(x)~-2\t a_0(x)G_2(x)\}\label{ham}
\ee
This gives the equations of motion for the $Z_\a$ as
\be
iD_0Z_\a~=~D_iD_iZ_\a~-~{2i\t\over{\cal D}}[\nabla Z^\dagger(D_iD_iZ)~+~
(D_iD_iZ)^\dagger\nabla Z]\times \nabla Z_\a
\ee
where $D_0\equiv\d_0-iA_0$. In the limit $\t\rightarrow 0$ we get the 
the equations given in \cite{chaks}. The equation for $a_k(x)$ is found to
be 
\be
\d_ta_k(x)~=~-{1\over {\cal D}}[\d_kZ^\dagger(D_iD_iZ)~+~(D_iD_iZ)^\dagger
\d_kZ]
\ee
The translation generator turns out to be 
\be
P_i=\int d^2x [{\cal P}_{\a}(x) \d_iz_{\a}(x)+{\cal P}^*_{\a}(x) \d_iz^*_{\a}(x)
+\pi_j(x)\d_i a_j(x)]\label{momentum}
\ee
This, when using the definitions of momenta reduces to 
\be
P_i~=~-{i\over 2}\int d^2x Z^\dagger {\stackrel{\leftrightarrow}{\d_i}}Z~-~
2\t\int d^2x a_i(x)G_2(x)\label{momentum1}
\ee
One can easily show after a straightforward calculation that 
\be
\{Z_\a(x),P_i\}^{DB}~=~\d_iZ_\a
\ee
Similarly one gets for $a_i$
\be
\{a_k(x),P_i\}~=~\d_ia_k.
\ee
At this stage it is instructive to calculate the $\left\{P_i,P_j\right\}^{DB}$.
We can just proceed as was done for the model without Hopf term to
find that
\be 
\left\{P_i,P_j\right\}^{DB}=2 \pi N \e_{ij}\label{ctrlcharge}
\ee
where $N= \int j_0(x)$ is topological charge given in eq.(\ref{soliton}).
Comparing the similar equation given in \cite{chaks}
we conclude that
presence of the Hopf term does not affect the central charge,
{\em viz.,} the right hand side of eq.(\ref{ctrlcharge}).

Now let us consider the angular momentum generator. It is given by
\be
J= \int d^2x~ [\e_{ij}x_ip_j + \pi_i \Sigma^{12}_{ij}a_j]\label{angmom}
\ee
where $p_i$ is the momentum density and is the integrand in
(\ref{momentum}):
\be
p_i= {\cal P}_{\a}(x) \d_iz_{\a}(x)+{\cal P}^*_{\a}(x) \d_iz^*_{\a}(x)
+\t \e_{jk}a_k(x)\d_i a_j(x)
\ee
and
\be
\Sigma^{12}_{ij}=\de_{1i} \de_{2j}- \de_{2i}\de_{1j}
\ee
We can write a more simplified form of angular momentum in 
eq.(\ref{angmom}) as 
\be
J= \int d^2x [\e_{ij}x_ip_j + \t a_ja_j]
\ee
Using this one can indeed show after a straightforward calculation
that appropriate spatial rotation is generated by $J$,
\be
\left\{Z(x),J\right\}=\e_{ij}x_i\d_j Z(x) 
\ee
\be\left\{a_k(x),J\right\}=\e_{ij}x_i\d_j a_k(x) + \e_{ki}a_i(x)
\ee
Inspite of the dramatic changes in the 
symplectic structure all these generators effect the same transformations as 
the model without the Hopf term. Ofcourse in the limit of
the coefficient of the Hopf term vanishing we get back the conventional 
model offering Landau-Lifshitz description of Ferromagnets. 
Note the ubiquitous presence of ${\cal D}$ in all the new Dirac brackets.
Further analysis can be done by considering quantum fluctuations in 
a skyrmion background. Eventhough
the spin algebra could not be reconstructed when $\t~\neq~ 0$ in a 
general analysis,
in a fixed background of soliton density profile we can obtain the
algebra through rescaling. Then one can consider fluctuations about
this background.
The new algebraic description arising here will play an important role
for the quantum fluctuations and will be presented elsewhere. 

\noindent{\bf Acknowledgements:}
TRG would like to thank Prof. A. P. Balachandran for discussions.
Part of the work was done at Dept of Phys, Syracuse University
supported by the DOE grant DE-FG02-85ER40231.

\end{document}